\documentclass{PoS}
\newcommand{\be}{\begin{equation}}
\newcommand{\ee}{\end{equation}}  	 
\newcommand{\ba}{\begin{eqnarray}}      
\newcommand{\ea}{\end{eqnarray}}	 
\newcommand{\bite}{\begin{itemize}}
\newcommand{\eite}{\end{itemize}}

\usepackage[vcentermath]{youngtab}
\usepackage{cite}

\title{Charmed Baryon Spectroscopy from 
lattice QCD for $N_f = 2+1$ flavours}

\ShortTitle{Charmed Baryon Spectroscopy}

\author{\speaker{Paula P\'erez Rubio}\\
        Institut f\"ur Theoretische Physik, Universit\"at Regensburg,
        D-93040 Regensburg, Germany.\\
        E-mail: \email{paula.perez-rubio@physik.uni-r.de}}

\abstract{ In recent years, several charmed baryons have been 
discovered, with more states likely to be found in the future. We investigate 
the spectra of singly and doubly charmed baryons on the lattice. 
The spin J=1/2 and J=3/2 states are calculated for both 
positive and negative parity.}

\FullConference{Xth Quark Confinement and the Hadron Spectrum,\\
		October 8-12, 2012\\
		TUM Campus Garching, Munich, Germany}

\begin{document}

\vspace{-0.4em}
\section{Introduction}

\vspace{-0.4em}
Baryons containing heavy quarks provide 
an interesting laboratory for studying QCD. They 
combine two different regimes: the slow relative motion of the 
heavy quarks with the relativistic motion of a light
quark.
The energies, detectors and luminosities at disposal in experiments 
made possible the observation of heavy baryons with
one heavy quark.
The first heavy  baryon signal
ever seen was the $\Lambda_c^+$ at BNL in 1975 followed 
soon after (1976) by the discovery of  $\Sigma^{++}_c$ at FNAL.
Also at Fermilab, in 1981, the first 
baryon containing a bottom quark $\Lambda_b$  was observed. 
Substantial progress has been made, 
and currently, there is a total 
of 19 charmed  and 5 bottomed baryons in the PDG  summary tables 
\cite{Beringer:1900zz}. In the last decade,  
charmed baryons have mainly  been  observed at the B-factories, 
whereas new bottom baryons were found at the Tevatron and more 
recently also at the LHC. Masses, decay form factors, lifetimes
and widths have been determined. However, 
identification of the spin and parity quantum numbers from 
experiments is still missing; so far, they are assigned based on quark
model expectations. It is expected that the large statistics 
provided by the LHC will allow for their identification
through the study of angular distributions of the particle decays.
Also, the PANDA experiment at the 
FAIR facility and the KEK Super-B Factory will look 
for charmed baryon signals.
For doubly heavy baryons, the experimental situation is 
less favourable. In  2003, Selex \cite{Mattson:2002vu} published evidence for 
$\Xi_{cc}^+$, but no other experiment has so far confirmed this channel. 

\vspace{0.4em}

Inspired by the experimental activity, 
many theoretical approaches have been used that try 
to  reproduce the existing spectra and predict 
new states: for example quark models~\cite{Copley:1979wj,Capstick:1986bm,
Roncaglia:1995az,SilvestreBrac:1996bg,Ebert:2002ig,Ebert:2005xj,
Roberts:2007ni,Garcilazo:2007eh,Valcarce:2008dr}, 
 QCD sum rules~\cite{Bagan:1991sc,Bagan:1992tp,
Wang:2002ts,Wang:2007sqa}, heavy quark effective theory 
(HQET) based models~\cite{Jenkins:1993ta}, 
and lattice QCD (LQCD)~\cite{Bowler:1996ws,Na:2007pv, 
Liu:2009jc, Alexandrou:2012xk, Briceno:2012wt,Durr:2012dw, Basak:2012py, 
Namekawa:2013vu}.

\vspace{0.4em}

In this work, we focus on the study  of 
singly and doubly charmed baryon low lying spectra 
including positive and negative parity states 
using  2+1 light dynamical flavours. This is achieved 
through the Monte Carlo evaluation of a
path integral in  discretised  
Euclidean space time with a lattice spacing $a$.  
Systematic errors are under control if the ultraviolet
cutoff, $a^{-1}$, is larger than the scales of the 
problem, the size of the box, $L$, is larger 
than the typical size of the hadrons under consideration 
and an extrapolation to the physical quark masses is performed.
Considering the hierarchy of the quark masses and energy scales,  
for  current lattice spacings, 
$$m_{\rm light} < m_s \sim \Lambda <m_c < 1/a < m_b$$
where $\Lambda$ is the typical hadronic scale. 
The charm quark mass is below the cutoff and a relativistic calculation 
of the baryon spectra containing charm quarks is viable. 
However, it is important to show discretisation errors 
are under control. 

\vspace{0.4em}

This write-up is organised as follows. In section 2, details of 
the computational setup  are given.  In section 3 we
briefly describe the methodology used to extract the 
energy levels. Following this, a detailed list of interpolating operators
is given and results are presented in section 4.
We finish with some concluding remarks in section 5.

\vspace{-0.7em}

\section{Computational details}

\vspace{-0.8em}
In our calculations, two different sets of gauge configurations with 
$u/d$ and $s$ sea quarks were used, 2-HEX~\cite{Durr:2010aw} 
and SLiNC~\cite{Bietenholz:2010jr,Bietenholz:2011qq}.
They were generated by the BMW-c and QCDSF collaborations, respectively. 
They were both generated using a gauge action where ${\rm O}(a^2)$
effects were reduced. 
The 2-HEX configurations employ tree-level clover Wilson fermions 
coupled to links with two steps of HEX smearing \cite{Capitani:2006ni}.  
This means that  the lattice artifacts are reduced to ${\rm O}(\alpha_sa)$.
SLiNC configurations use non-perturbatively clover Wilson fermions 
with stout link smearing on the derivative terms\cite{Cundy:2009yy}:
${\rm O}(a)$ effects are non-perturbatively removed.

\vspace{0.5em}
The 2-HEX ensembles were generated for a range of lattice  spacings 
from  $a \sim 0.092$~fm down  to $a\sim 0.054$~fm and pion masses from 
$M_{\pi} \sim 520$~MeV down to $M_\pi \sim 120$~MeV.
The spatial dimensions cover the range   $1.7 -5.9$~fm 
and  the number of configurations per ensemble used for this study
 is $\sim 200$. See \cite{Durr:2010aw} for more details. With this 
set of ensembles  we can perform a  controlled chiral and continuum 
extrapolation. 

\vspace{0.5em}
For the SLiNC configurations, there is only one lattice spacing available, 
$a\sim 0.0795$~fm. The light quark masses 
were  tuned to the 
SU(3)$_{\rm flavour}$-symmetric  point, where the flavour singlet 
mass average $m_q = (m_u + m_d + m_s)/3$ takes its physical value. 
Then, $m_{u,d}$ and $m_s$ are varied keeping $m_q$ constant 
\cite{Bietenholz:2010jr,Bietenholz:2011qq}. 
This is motivated by the Gell-Mann - Oakes - Renner relation
as well as by SU(3) chiral perturbation theory ($\chi$-PT):
$m_K$ approaches its 
physical value from below, so that it is  within the range of 
applicability of  $\chi$-PT ($m_{K} < 600$~MeV).  
Ensembles are available for two spatial extents, 
$1.9$~fm and $2.5$~fm. So far, we have analysed one of the  
smaller configurations, with $M_{\pi}=348$ MeV
and $M_K=483$ MeV.

\vspace{-0.6em}
\section{Methodology}

\vspace{-0.6em}
Let $\hat \mathcal O_1$, $\hat \mathcal O_2$ be two interpolating operators 
that overlap with the baryonic state we are interested in. 
The correlation function of $\hat \mathcal O_1$ and 
$\hat \mathcal O_2$ is given by, 
\ba
C_{{\bf p}=0}(\hat  \mathcal O_1,\hat \mathcal O_2,t) & = &
\langle \hat \mathcal O_1(0) \overline{ \hat \mathcal { O}_2} (t)  \rangle = 
\lim_{T\to \infty} \frac{1}{Z(T)} {\rm Tr} \left[e^{-(T -t)\hat H} 
\hat \mathcal O_1 e^{-t\hat H}\overline{\hat \mathcal O_2}\right] \nonumber \\
&=& \sum_n \langle 0| \hat \mathcal O_1 |n \rangle  
\langle n| \overline{\hat \mathcal O_2}|0 \rangle e^{-E_nt}, \quad 
Z(T) = {\rm Tr}\left[e^{-T\hat H}\right], 
\ea
where $T$ is the temporal extent of the lattice.  Every correlation 
function contains contributions from all states with the same quantum numbers. 
Since we are interested in the low lying spectra, it is desirable to 
eliminate the contribution of excited states.
This can be done using the variational method \cite{Michael:1985ne,
Luscher:1990ck}.  

\vspace{-0.5em}
\subsection{Variational method}

The idea is to choose a basis of operators $\hat \mathcal O_i$  with different overlaps
with the state we are interested in. A suitable possibility  is to create 
operators with the same Fock structure and different spatial extent. This can be
achieved by applying different numbers of steps of  Wuppertal smearing 
\cite{Gusken:1989ad,Gusken:1989qx} to the fermionic fields.  A cross correlation
matrix can then be built, $\left[ C(t)\right]_{ij} = C(\hat\mathcal O_i,
\hat\mathcal O_j,t)$. Solving the generalised eigenvalue problem (GEVP), 
\be
C^{-1/2}(t_0) C(t) C^{-1/2}(t_0) v^\alpha(t,t_0) = \lambda^\alpha(t,t_0) 
v^\alpha(t,t_0).
\ee
It can be shown that the eigenvalues have the following 
behaviour, 
\be
\lambda^\alpha(t,t_0) \propto  e^{-E_\alpha(t-t_0)} 
\left[ 1 + {\rm O}\left(e^{-\Delta E_\alpha t} \right) \right],
\quad \Delta E_\alpha  = E_{\alpha'} - E_\alpha, \hspace{1em }\mathrm{ with } 
\hspace{1em}\alpha' > \alpha.
\ee
In this way, the lower levels can be extracted 
cleanly. Nevertheless, there is some freedom of choice of the 
structure of the interpolating operators. We now motivate the 
choices made in this work.

\subsection{Interpolating operators}

To construct sensible interpolating 
operators overlapping with the baryon states, it is
useful to think in terms of their possible inner structure. 
The constituent quark model guided by the approximate SU(2) or SU(3) 
flavour symmetry is quite successful when applied 
to the light quark sector. This statement may no longer be
true when we consider heavy baryons. As far as singly 
charmed baryons are concerned, predictions of HQET
tell us  that the light degrees of freedom form a diquark and move 
around the  approximately static heavy colour source. 
In the limit $m_{Q} \to \infty$, the total angular momentum
of the light diquark becomes a good quantum number. Its total 
spin $s_d$ takes two possible values, $0$ and $1$ corresponding to
a flavour antisymmetric or  symmetric structure, respectively.
It is possible to construct operators following these prescription, 
cf. Table \ref{label3}. 
\begin{table}[ht!]
{\footnotesize
\begin{center}
\setlength{\tabcolsep}{0.7mm}
\renewcommand{\arraystretch}{1.2}
\begin{tabular}{|rcr|c|c|c|c|}
\multicolumn{5}{c}{{\sc Singly charmed }} 
&\multicolumn{1}{c}{{$J= \frac12$}}
 &\multicolumn{1}{c}{{$J= \frac32$}}\\ 
\hline
  $\bf (S)$& $\bf (I)$& $\bf s_d$ & $\bf (qq)Q$ & $\mathcal O$  &\textbf{Name}&\textbf{Name}\\
\hline
\hline
 $(0)$ & $(0)$                  & $(0)^+$& $(ll')c$ & 
 ${\mathcal O}_5=\epsilon_{abc}(l^{aT}C\gamma_5l^{'b})c^c  $&$\Lambda_c$ &  \\
 $(0)$ & $(1)$                  & $(1)^+$& $(ll)c$ &
 ${\mathcal O}_\mu=\epsilon_{abc}(l^{aT}C\gamma_\mu l^b)c^c $&$\Sigma_c$  &$\Sigma^*_c$   \\
 $(-1)$& $\left(\frac12\right)$ & $(0)^+$& $(ls)c$ &
 ${\mathcal O}_5=\epsilon_{abc}(l^{aT}C\gamma_5 s^b)c^c $&$\Xi_c$     &      \\
 $(-1)$& $\left(\frac12\right)$ & $(1)^+$& $(ls)c$ &
 ${\mathcal O}'_\mu=\epsilon_{abc}(l^{aT}C\gamma_\mu s^b)c^c $&$\Xi'_c$    &$\Xi^*_c$     \\
 $(-2)$& $(0)$                  & $(1)^+$& $(ss)c$ &
 ${\mathcal O}_\mu=\epsilon_{abc}(s^{aT}C\gamma_\mu s^b)c^c $&$\Omega_c$  &$\Omega^*_c$   \\
\hline
\hline
\end{tabular}
\caption{\label{label3} \footnotesize 
Summary of quantum numbers of singly charmed baryons 
and interpolating operators following  the HQET approach. 
$l$ and $l'$ stand for light quarks and 
$S,I$ and $s_d $ stand for
strangeness, isospin, and diquark total spin, respectively.  
These operators were  suggested in 
\cite{Bowler:1996ws} for lattice calculations.
}
\vspace{-1.7em}
\end{center}
}
\end{table}

As we can see, operators $\mathcal O_\mu$ (with $s_d=1$)  have contributions from 
states with total spin $\frac 12$ and $\frac 32$. We thus have to 
use spin projectors to disentangle them. For zero momentum, these 
projections amount to 
\vspace{-0.5em}
\ba
(P^{3/2})_{ij} &=& \delta^{ij} -\frac{1}{3} \gamma^i\gamma^j, 
\quad i,j= \{1,2,3\}, \nonumber \\ 
(P^{1/2})_{ij} &=&  \frac{1}{3} \gamma^{i}\gamma^{j}.
\ea

Since flavour symmetry is not a good symmetry for 
charmed baryon systems, in principle it is not clear that 
interpolating operators falling into the irreducible representations 
of SU(4)$_{\rm flavour}$ have a decent overlap with the 
baryon states. However, these states have also been constructed, 
and we have seen, cf. section 4, that they lead to results 
that are compatible with the energy levels obtained from 
the HQET based interpolating operators. The
tensor product of three fundamental representations of
SU(4)$_{\rm flavour}$ is, 
 \ba
 \hspace{1.1mm}4\hspace{1.1mm}\otimes \hspace{1.1mm}4\hspace{1.1mm} \otimes
 \hspace{1.1mm}4 \hspace{1.1mm}&=& \hspace{4mm}20_{\rm S}\hspace{3.5mm} \oplus
 \hspace{0.4mm} 20_{\rm M}\hspace{0.5mm} \otimes \hspace{0.4mm}20_{\rm M}
 \hspace{0.4mm}\otimes\hspace{1.4mm} \overline 4\hspace{0.3mm} \nonumber\\
 \yng(1)\otimes \yng(1) \otimes \yng(1)& =& \yng(3) \oplus \yng(2,1) \otimes
 \yng(2,1) \otimes \yng(1,1,1)\nonumber
 \ea
We can construct interpolating operators falling into these multiplets. 
Since there are at most three different flavours in a baryon, the structure 
is exactly the same as for the interpolating operators in the SU(3)$_{\rm flavour}$
multiplets (with $u,d,s$ quarks). We list the operators 
corresponding to  baryons containing $(ll^{(')}c)$ as constituents.

 \bite
 \item{ \footnotesize {\sc SU(4) 20-plet containing  SU(3) octets
 $\yng(2,1)$}
 \vspace{1mm}
 \bite
 \item[]{{\bf $\Sigma_c$}: \hspace{1em} 
     $\mathcal O^{P}_\gamma(x)=\epsilon^{abc} 
       \left[ c^a(x)^{T}(C\gamma_5)l^b(x)
       \right]l^c_{\gamma}(x). \hspace{4.7cm}$ }
 \item[]{{\bf $\Lambda_c$ }:\hspace{1em}
     \scriptsize {$  \mathcal O^{\Lambda}_\gamma(x)=\frac{1}{\sqrt{6}}\epsilon^{abc} 
       \left\{ 2 \left[ l^a(x)^{T}(C\gamma_5)l^{'b}_2(x)\right] 
         c^c_{\gamma}(x)  +
         \left[ c^a(x)^{T}(C\gamma_5)l^{'b}(x)\right]
         l^c_{\gamma}(x) - \left[ c^a(x)^{T}(C\gamma_5)l^b(x)\right]
          l^{'c}_{\gamma}(x)  
         \right\}.$} }
 \item[]{{\bf $\Sigma^0_c$}:  \hspace{1em}
     $\mathcal O^{\Sigma_0}_\gamma(x)=\frac{1}{\sqrt{2}}\epsilon^{abc} 
       \left\{  \left[ l^a(x)^{T}(C\gamma_5)c^b(x)\right] 
         l^{'c}_{\gamma}(x)  +
         \left[ l^{'a}(x)^{T}(C\gamma_5)c^b(x)\right]
         l^c_{\gamma}(x) \right\}.$
 }
 \eite
}
 \item{\footnotesize {\sc SU(4) 20-plet containing SU(3) decuplet $\yng(3)$}
 \vspace{1mm}
 \bite
 \item[]{{\bf $\Sigma_c^{*}$ }:\hspace{1em} 
  {$\mathcal O_{\gamma}^{\Sigma^{*-}} = \epsilon^{abc}\left\{ 
 2\left( c^{aT}(C\gamma_\mu)l ^b \right)l^c_{\gamma} +
 \left( l^{aT}(C\gamma_\mu)l ^b \right)c^c_{\gamma} \right\}.$}
 }
 \item[]{{\bf $\Sigma^{*0}_c$}: \hspace{1em}
 \scriptsize{$\mathcal O_{\gamma}^{\Sigma^{*0}} =\frac{\epsilon^{abc}}{\sqrt 3}\left\{
 \left( l^{aT}(C\gamma_\mu)l^{'b} \right)c^c_{\gamma}+ 
 \left( c^{aT}(C\gamma_\mu)l^b \right)l^{'c}_{\gamma}+ 
 \left( l^{'aT}(C\gamma_\mu)c^b \right)l^c_{\gamma}\right\}. 
 $}
 }
 \eite
 }
\eite

The operators for the baryons with strange content are obtained 
by substituting the $l$ quark by $s$ in the operators given above.

\begin{figure}[h]
\begin{center}
\vspace{0.3em}
\begin{minipage}{14pc}
\includegraphics[width=7pc]{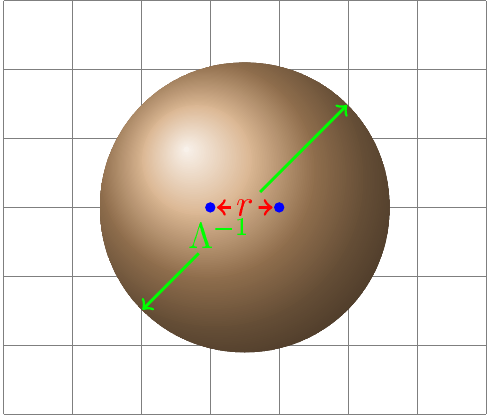}
\end{minipage}\hspace{2pc}%
\begin{minipage}{14pc}
\includegraphics[width=7pc]{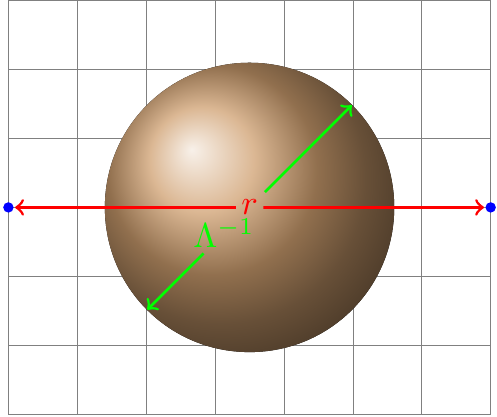}
\end{minipage} 
\caption{\label{label1}  \footnotesize{ Structure 
of a doubly heavy baryon: HQET picture (left), 
Quarkonium-like picture (right), }}
\vspace{-1.5em}
\end{center}
\end{figure}

Next, we  discuss possible interpolating operators for the 
doubly charmed baryons.  
 In Figure \ref{label1}, 
we depict two possible structures for these systems. 
 On the 
left hand side, the HQET picture, the diquark is formed by 
the two heavy quarks $QQ$ and interacts with the 
remaining light quark as if it was a heavy light meson. 
In this case, the radius of the $QQ$ system is much 
smaller than $\Lambda^{-1}$.  
The operators reflecting this structure are listed 
in Table \ref{label4}.

\begin{table}[ht!]
{\footnotesize
\begin{center}
\setlength{\tabcolsep}{0.7mm}
\renewcommand{\arraystretch}{1.2}
\begin{tabular}{|rcr|c|c|c|c|}
\multicolumn{5}{c}{\sc  Doubly charmed}&\multicolumn{1}{c}{{$J= \frac12$}}
 &\multicolumn{1}{c}{{$J= \frac32$}}\\ 
\hline
  $\bf (S)$& $\bf (I)$& $\bf s_d$ & $\bf (QQ)q$ & $\mathcal O$  &\textbf{Name}&\textbf{Name}\\
\hline
\hline
 $(0)$ & $(0)$                  & $(1)$& $(cc)l$ & ${\mathcal O}_\mu=\epsilon_{abc}(c^{aT}C\gamma_\mu c^b)l^c $&$\Xi_{cc}$    &$\Xi^*_{cc}$     \\
 $(-1)$& $\left(\frac12\right)$ & $(1)$& $(cc)s$ & ${\mathcal O}_\mu=\epsilon_{abc}(c^{aT}C\gamma_\mu c^b)s^c $&$\Omega_{cc}$ &$\Omega^*_{cc}$  \\
\hline
\end{tabular}
\caption{\label{label4} \footnotesize Summary of quantum numbers of doubly heavy baryons 
and interpolating operators following  the HQET approach. 
$l,l',S,I$ and $s_d $  are the same as in Table \protect\ref{label3}.}
\vspace{-1.7em}
\end{center}
}
\end{table}
On the right hand side, the quarkonium-like picture, the diquark 
is formed by a heavy and a light quark ($Qq$) and will 
interact with the remaining $Q$ as if it was a $\bar Q$. 
In this case the radius of the $QQ$ system is larger 
than $\Lambda^{-1}$. An example of these   operators
is given by 
 \bite
 \item[]{{\bf $\Xi_{cc}$}: \hspace{1em} 
     $\mathcal O^{P}_\gamma(x)=\epsilon^{abc} 
       \left[ l^a(x)^{T}(C\gamma_5)c^b(x)
       \right]c^c_{\gamma}(x). \hspace{4.7cm}$. }
\eite

Once the 
correlation functions are constructed out of two interpolating 
operators, one has to project the correlator into the desired 
parity, i.e.,
\be
C(\hat \mathcal O_i, \hat \mathcal O_j,t) = 
T_{\bar \gamma\gamma} \langle \hat \mathcal O_{i,\gamma}(0)
\overline{\hat \mathcal O_{j,\bar \gamma}}(t)\rangle, 
\ee
where $T$ is a polarisation matrix that projects onto positive 
or negative parity.

\vspace{-0.8em}
\section{Results}

\vspace{-0.6em}
\subsection{Continuum limit}

\begin{center}
\begin{figure}[h]
\begin{minipage}{17pc}
\includegraphics[width=15pc]{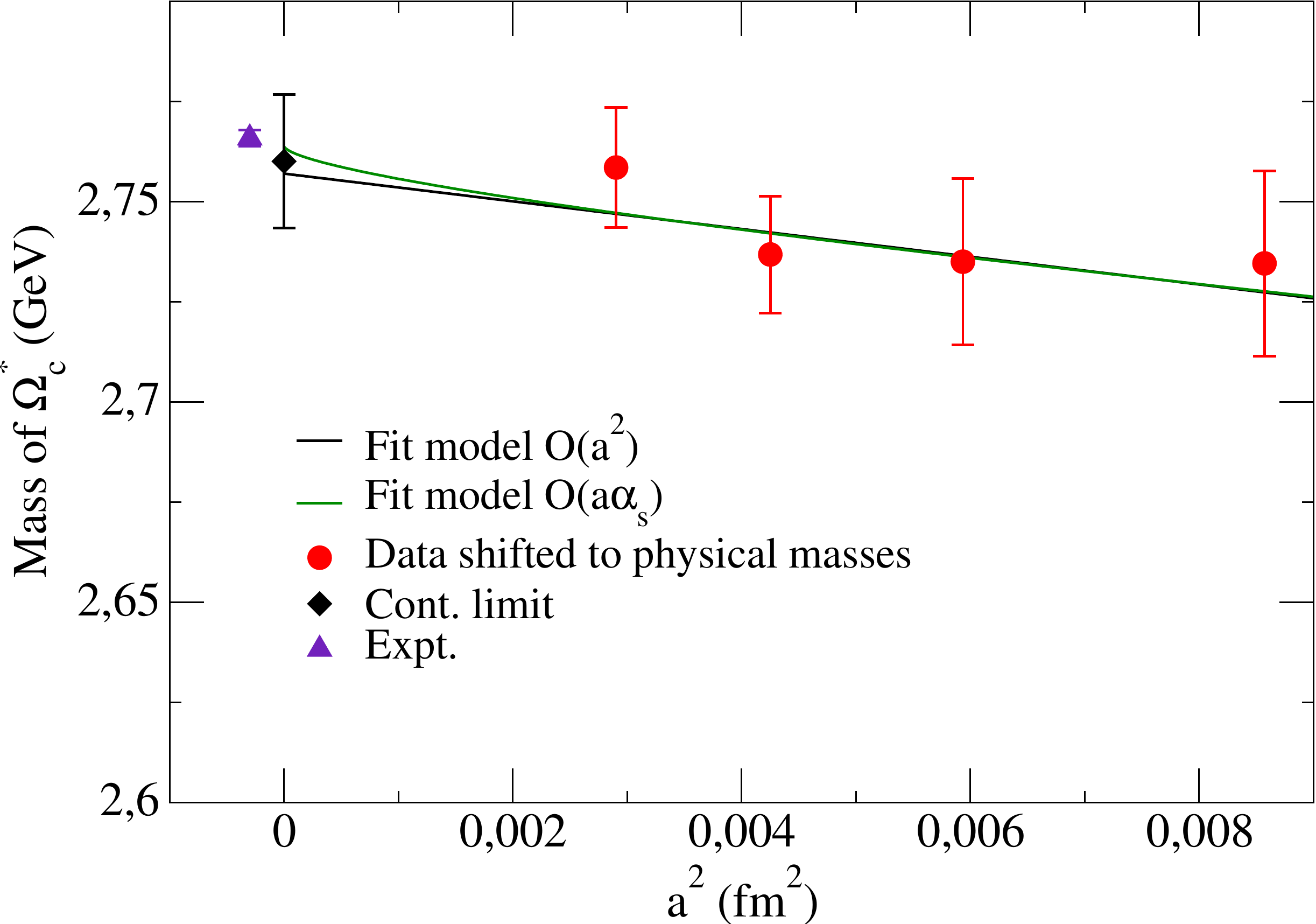}
\end{minipage}\hspace{2pc}%
\begin{minipage}{17pc}
\includegraphics[width=15pc]{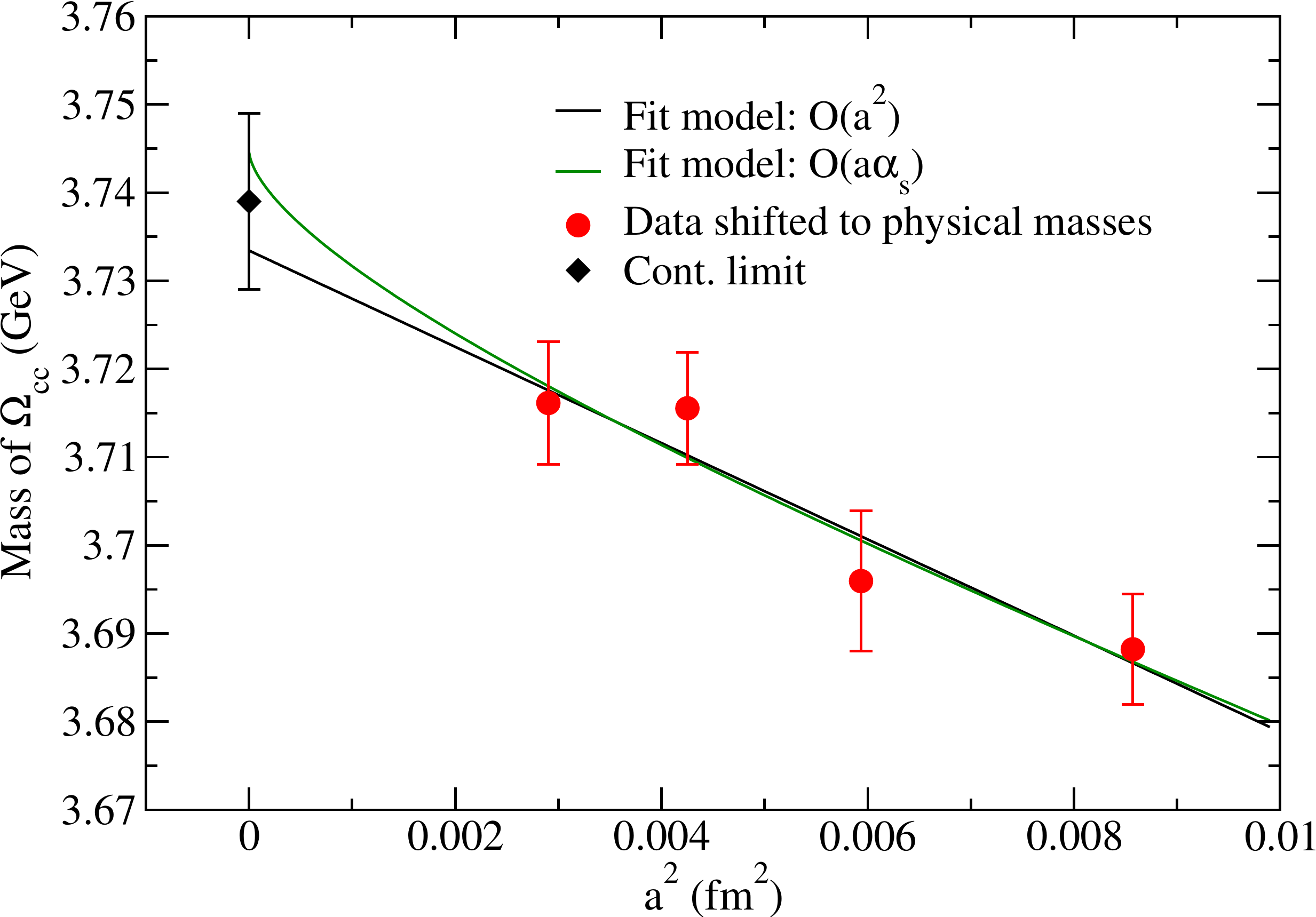}
\end{minipage}
\vspace{-1pc} 
\caption{\label{label6}\footnotesize{{\sc Preliminary}.
Continuum limit extrapolation from the 
2-HEX configurations for $\Omega^*_c$ (left)
and $\Omega_{cc}$ (right) states. The Triangle (violet)
stands for the experimental value. The diamonds (black) correspond
to the continuum extrapolation and the circles (red) are the 
simulation data  at finite lattice spacing shifted to the physical 
masses of the light quarks. Two fitting functions have been used, 
${\rm O}(a^2)$ and ${\rm O}(\alpha_sa)$.
}}
\vspace{-0.5em}
    \end{figure}
\end{center}

In Figure \ref{label6},
we present preliminary results for the continuum 
limit extrapolation of the $\Omega^*_c$ and 
the $\Omega_{cc}$ using results from the 
2-HEX configurations. 
A combined fit including the lattice 
spacing and the light quark mass dependences
has been carried out. The following fit function 
to the baryon masses has been used,
\vspace{-0.6em}
\be
y^{\rm FIT} (a, M_\pi, M_{\bar ss}|{\bf A}) = 
y^{\rm cont} f_X(a)\left(1 + A_2 x_\pi\right)
\left(1 + A_3 x_s\right),
\ee
where $M_{\bar ss}$ refers to the strange-strange pseudoscalar and 
\vspace{-0.6em}
\be
x_\pi = \frac{(M_\pi^{\rm latt})^2 - (M_\pi^{\rm phys})^2}{(M_\pi^{\rm phys})^2},
\quad
x_s = \frac{(M_{\bar ss}^{\rm latt})^2 - (M_{\bar ss}^{\rm phys})^2}{(M_{\bar ss}^{\rm phys})^2}.
\ee
For the continuum extrapolation, two fit functions have been used, 
\vspace{-0.6em}
\be
f_1(a) = 1 + A_1 a^2 , \quad f_2(a) = 1 + A_2 a\alpha_s.
\ee
$f_1$ is also considered since one expects one loop corrections to 
be small due to the use of smeared links in the action. The errors were
estimated using the histogram method 
combined with the bootstrap statistical analysis 
\cite{Durr:2010aw,Durr:2008zz,Durr:2011ap}.
Different fit ranges  were chosen and only those with 
$\chi^2/{\rm dof} < 2$ were included. 

\vspace{0.5em}
In Figure \ref{label6} the results for the masses 
of $\Omega_c^*$ (left) and $\Omega_{cc}$  (right)
are shown. To illustrate the quality of the fit, 
the data points are shifted to the physical 
light quark mass values and then averaged 
for each lattice spacing. 

\vspace{-1.5em}

\vspace{-2.7em}
\subsection{Spectra}

\begin{center}
\begin{figure}[h]
\vspace{-1.0em}
\begin{minipage}{17pc}
\includegraphics[width=16pc]{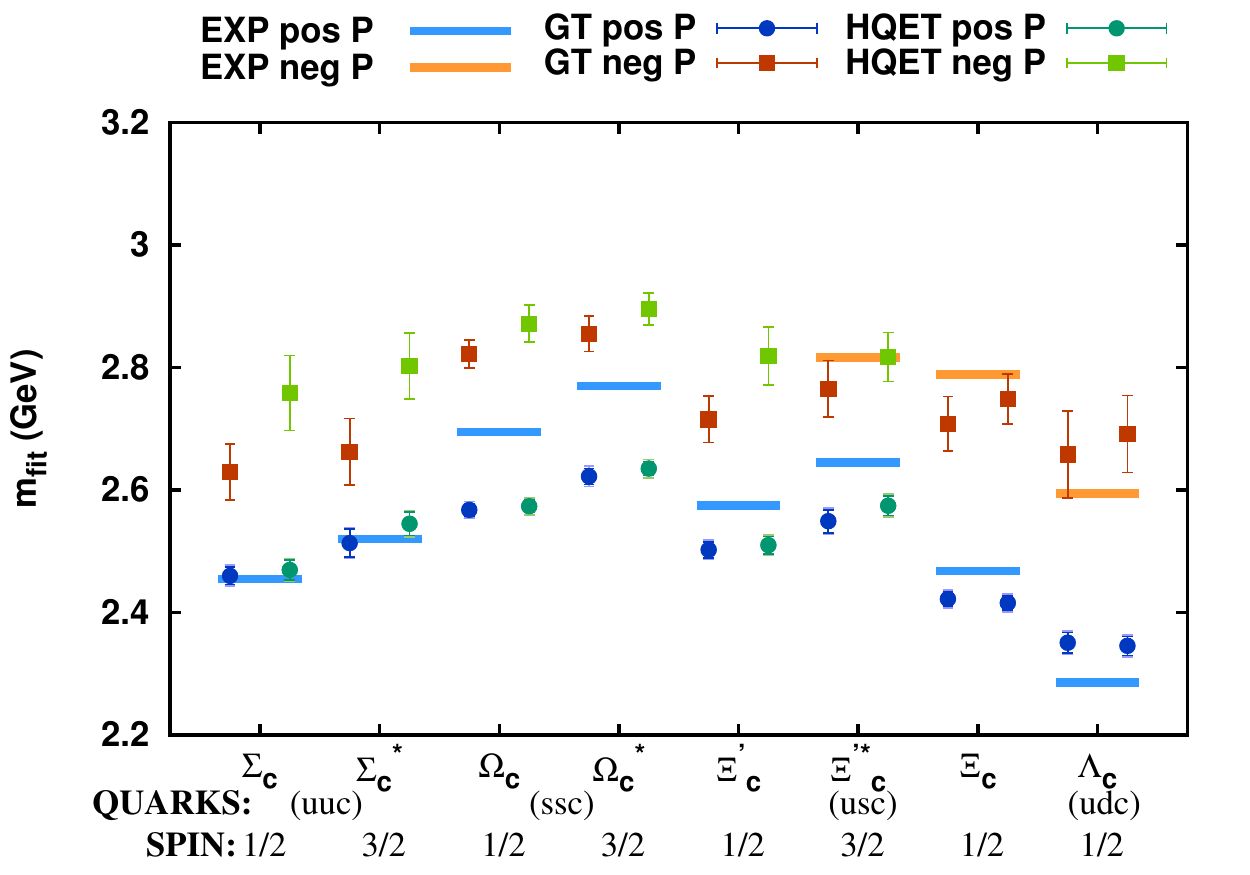}
\end{minipage}\hspace{2pc}%
\begin{minipage}{17pc}
\includegraphics[width=16pc]{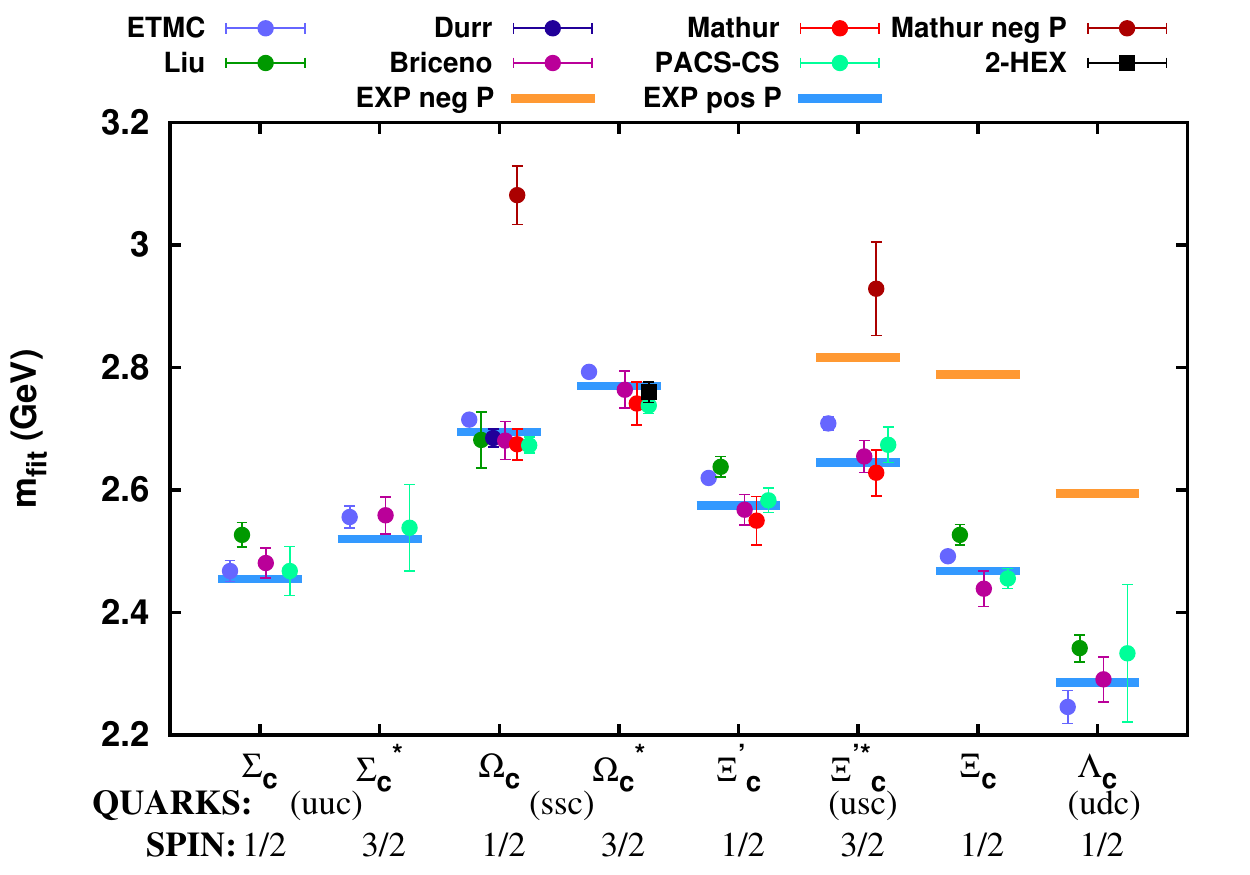}
\end{minipage}
%
\begin{minipage}{17pc}
\includegraphics[width=16pc]{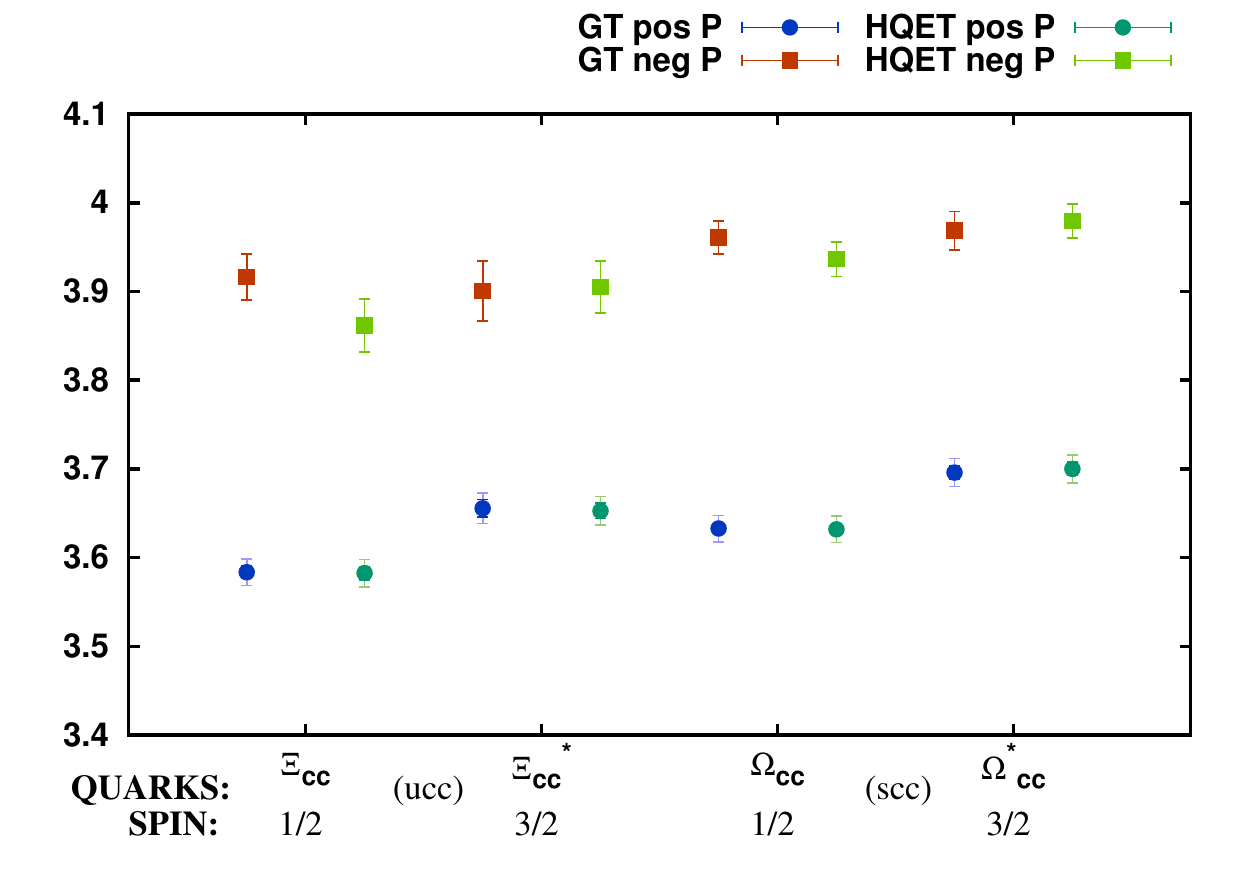}
\end{minipage}\hspace{2pc}%
\begin{minipage}{17pc}
\includegraphics[width=16pc]{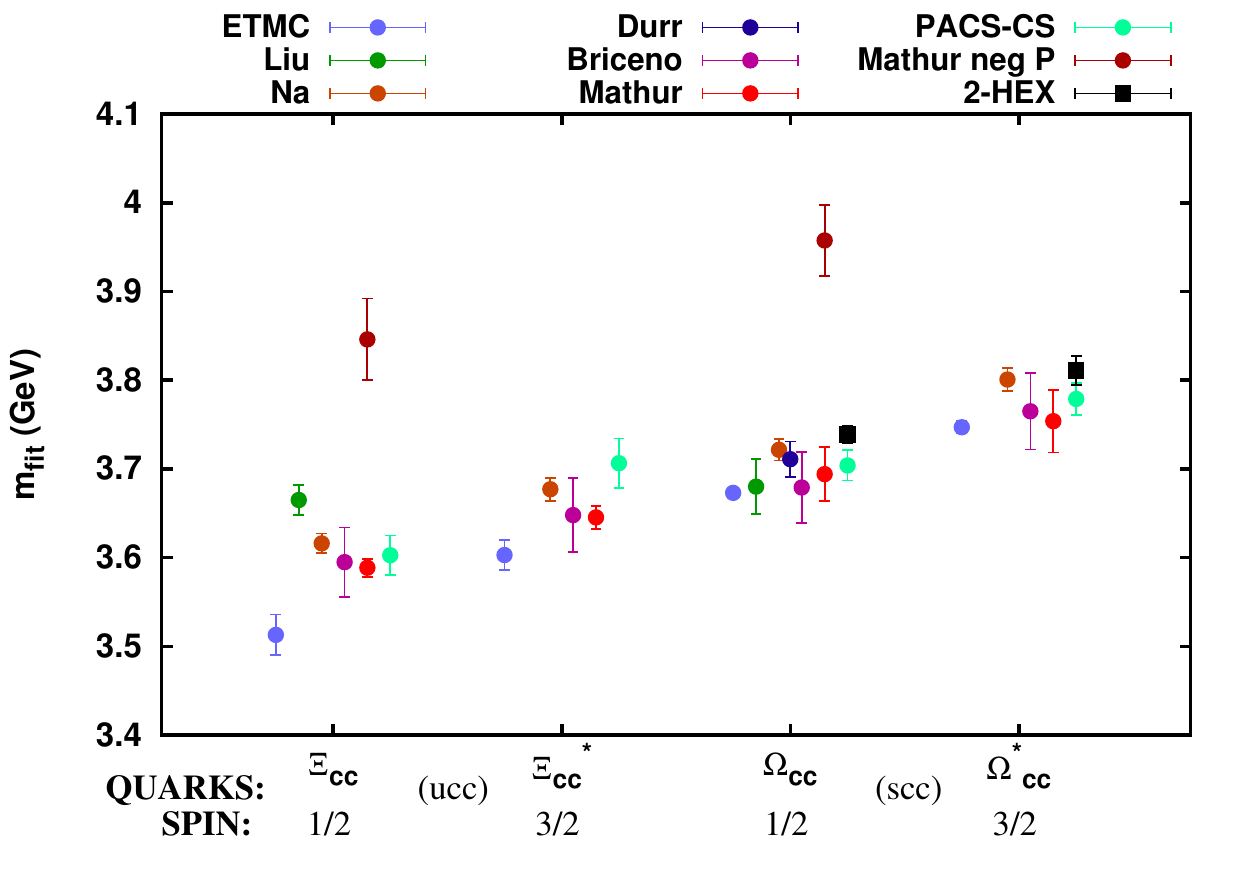}
\end{minipage} 
\caption{\label{label7}  \footnotesize{
Singly charmed (top) and doubly charmed (bottom) low 
lying spectrum. On the left hand side, results from the
SLiNC configurations are shown ($M_\pi = 348 $MeV). 
The errors are statistical
only. On the right hand side, a summary of 
lattice results is presented, including results from 
this work (2-HEX, black squares). 
}}
\vspace{-1.8em}
    \end{figure}
\end{center}

The  charmed baryon spectra for the 
SLiNC ensemble are shown in Figure 
\ref{label7} (left). A
 $3\times 3$ correlation matrix has 
been constructed per interpolating operator, with three levels of 
smearing. 
We show results for the 
low lying singly (above)
and doubly (below) charmed baryon spectra, including
negative parity states, compared with the  experimental results. 
Simulations are still at an early 
stage and only one combination of light quark masses has been 
analysed. We can see that the mass differences between baryons
containing $u, d $ quarks and the ones with $s$ quarks 
are smaller than the experimental values.
This is not surprising  since we are far from the physical 
quark masses: the singlet quark mass $m_q$ is tuned to the 
physical value which means that the $u, d$ and $s$
quark masses are heavier and lighter 
than their physical values respectively, in spite of 
the fact that not all systematics are accounted for at present.
 On the right hand side, we can see a
summary of lattice results for the singly (above) and 
doubly (below) charmed spectra, with different 
systematics in each case~\cite{Na:2007pv, 
Liu:2009jc, Alexandrou:2012xk, Briceno:2012wt,Durr:2012dw, Basak:2012py, 
Namekawa:2013vu}. Also, the continuum extrapolated points from the 2-HEX 
configurations  are included. We can see that, overall, lattice results agree
with experiment.

\vspace{-2.9em}
\section{Conclusions and outlook}

\vspace{-0.6em}
Heavy baryons  are good systems to probe QCD dynamics. 
The last decade witnessed a huge experimental 
progress in the discovery of new singly heavy 
baryons. In the near future, spin and parity 
quantum number identification will be possible, 
thanks to the large statistics and advanced detectors 
at the LHC. In the longer term, further progress is 
expected from planned experiments (PANDA at the 
FAIR facility and KEK Super-B factory).

\vspace{0.4em}
In this write-up we have presented  preliminary 
results of an  on-going project to obtain  
the low lying spectra of singly and doubly charmed 
baryons on the lattice, including states with positive
and negative parity. Two different configuration 
ensembles are being used, employing SLiNC and 2-HEX 
fermions. 
 
\vspace{0.4em}

In the near future, we expect to analyse more sets of 
SLiNC configurations at different quark masses
and for larger volumes. For the 2-HEX 
ensembles we will focus on expanding the analysis to other states.

\vspace{-0.8em}
\section{Acknowledgements}
\vspace{-0.3em}
The numerical calculations were performed on the SGI Altix ICE machines at 
HLRN (Berlin-Hannover, Germany),
and the BlueGene/P (JuGene) and the 
Nehalem Cluster (JuRoPA) of the 
J\"ulich Supercomputer Center
and  the iDataCool cluster 
at Regensburg University. 
The Chroma software package 
\cite{Edwards:2004sx} was used for some of the analysis. 
I thank my colleagues from the QCDSF and BMW-c collaborations.
This work was supported 
by the EU ITN STRONGnet and the DFG SFB/TR 55.

\vspace{-0.8em}

\end{document}